\documentclass[twocolumn]{svjour3}
\RequirePackage[T1]{fontenc}
\smartqed

\RequirePackage{graphicx}
\RequirePackage{mathptmx}      
\RequirePackage{flushend}
\RequirePackage[numbers,sort&compress]{natbib}
\RequirePackage[colorlinks,citecolor=blue,urlcolor=blue,linkcolor=blue]{hyperref}

\RequirePackage{graphicx}
\RequirePackage{color}
\RequirePackage{multirow}
\RequirePackage{amsmath}
\RequirePackage{array}
\RequirePackage{booktabs}
\RequirePackage{multirow}
\RequirePackage{siunitx}

\newcommand{\be}{\begin{equation}}
\newcommand{\ee}{\end{equation}}
\newcommand{\bea}{\begin{eqnarray}}
\newcommand{\eea}{\end{eqnarray}}
\def\c#1{~\cite{#1}}
\def\f#1{Fig.~\ref{#1}}

\def\beq{\begin{equation}}
\def\eeq{\end{equation}}

\begin{document}
\title{Postponing  the dynamical transition density using competing interactions}
\author{Patrick Charbonneau \and
        Joyjit Kundu
}


\institute{Patrick Charbonneau \at Department of Chemistry, Duke University, Durham, North Carolina 27708, USA \\
\email{patrick.charbonneau@duke.edu}
          \and
          Joyjit Kundu \at Department of Physics, Duke University, Durham, North Carolina 27708, USA \\
          \email{joyjitkundu032@gmail.com}
          }

\date{Received: date / Accepted: date}



\maketitle

\begin{abstract}
Systems of dense spheres interacting through very short-ranged attraction are known from theory, simulations and colloidal experiments to exhibit dynamical {\em reentrance}. The liquid state can thus be fluidized to higher densities than otherwise possible with interactions that are purely repulsive or long-ranged attractive. A recent mean-field, infinite-dimensional calculation predicts that the dynamical arrest of the fluid can be further delayed by adding a longer-ranged repulsive contribution to the short-ranged attraction. We examine this proposal by performing extensive numerical simulations in a three-dimensional system. We first find the short-ranged attraction parameters necessary to achieve the densest liquid state, and then explore the parameters space for an additional longer-ranged repulsion that could enhance the effect. In the family of systems studied, no significant (within numerical accuracy) delay of the dynamical arrest is observed beyond what is already achieved by the short-ranged attraction. Possible explanations are discussed.
\end{abstract}

\section{Introduction} Particles with short-ranged attractive and long-ranged repulsive (SALR) interactions can form fairly elaborate structures\c{stradner_nat,bolhuis2014,yuan2016,ciach2013,truskett2016,bollingerI,bollingerII,truskett2017,cao:2017}. Despite the spherical symmetry of their pair interaction potential, at low temperatures these models assemble into exotic ordered and disordered mesophases, and their structural richness has clear dynamical consequences, even in the disordered regime\c{yuan2017,delgado2004,delgado_prl}. 
A recent theoretical proposal suggests that certain SALR models exhibit unusual dynamical features in the very dense fluid regime~\cite{francesco2018} as well. Maimbourg \emph{et al.}~\cite{francesco2018}'s extension of a high-dimensional treatment of the glass transition~\cite{patrick_review,sellitto2013,sellitto_epl} suggests that certain models should display a very pronounced dynamical reentrance upon changing temperature. More precisely, the theoretical analysis suggests that a carefully chosen high-density SALR system that is glassy at low temperature should, upon heating, first melt and then get dynamically arrested once again, all while remaining completely disordered, i.e., without crystallizing. 

On its own, such reentrance is not exceptional. The phase behavior of systems with core-softened interactions can exhibit multiple dynamically arrested phases leading to high-order singularities, as first proposed by mode-coupling theory \c{fabbian1999,fuchs1999,dawson2000,sperl2002,chen_soft_mater}, and then verified by both experiments \c{pham2002,chen2003,eckert2002,chen2000,lu2008,chen_soft_mater} and numerical simulations\c{gnan2014,sciortino2003,sciortino_pre,angel_jcp,angel2007,cates_pre,cates_prl,reichman2007}.  Dynamical quantities, such as the density-density correlator, then exhibit a logarithmic decay instead of a typical two-step relaxation, and the mean-squared displacement grows sub-diffusively instead of plateauing at intermediate times.
A common physical interpretation of this effect is that introducing short-ranged attraction leads to an interplay between two localization mechanisms: caging from the hard-core repulsion and interparticle bonding. As a result liquids with a higher packing fraction than is possible from either mechanism can then be stabilized~\cite{sciortino2002}. Adding a supplementary, longer-ranged repulsion is understood as effectively deepening the well created by the short-range attraction, and thus leads to a slightly more efficient packing of neighboring spheres in the liquid state\c{francesco2018}. In the mean-field description, the nonergodicity transition to a glass phase is then pushed to even higher densities although only over a very narrow temperature window\c{francesco2018}. Even though this improvement over a system with purely short-ranged attraction is predicted to be about $3\%$ in the $d\rightarrow\infty$ limit, the effect should be large enough to be numerically distinguishable if it indeed persists in a finite-dimensional system. An additional methodological challenge, however, is that this transition is only a crossover away from the $d\rightarrow\infty$ limit \c{patrick_review}. 

In this article, we attempt to test this prediction in three dimensions via extensive numerical simulations. First, we tune the attraction range of a system of particles interacting via a hard core followed by a short-ranged square-well attraction (SW) to maximize the high-density extension of the liquid phase. We then optimize the interaction parameters of a system with an additional larger-ranged square-shoulder repulsion (SW+SS) in an attempt to push the dynamical arrest to even higher densities. The plan for the rest of this article is as follows. In Sec.~\ref{sec:method} we describe the model, the simulation method and the observables of interest. In Sec.~\ref{sec:results}, we present the simulation results, and  we briefly conclude in Sec.~\ref{sec:concl}. 

\section{Models and Simulation Method} 
\label{sec:method}
We study $50\%-50\%$ binary (A-B) mixtures of $N=1000$ spherical particles interacting via two potentials: (i) a simple square-well (SW) interaction, and (ii) a SALR square-well plus square-shoulder (SW+SS) interaction. The hard core diameter ratio of the two particle types, $\sigma_{\rm A}/\sigma_{\rm B}=1.2$, with an additive hard-core interaction, i.e., $\sigma_{ij}=(\sigma_{i}+\sigma_{j})/2$ $\forall ij$, is chosen so as to strongly suppress crystallization. The interaction potential can then be generically expressed as
\begin{equation}
V_{ij}=
\begin{cases}\infty &r_{ij}\leq \sigma_{ij}\\ 
-U_0  &\sigma_{ij}<r_{ij}<\sigma_{ij}+\Delta^0_{ij}\\ 
U_1  &\sigma_{ij}+\Delta^0_{ij}<r_{ij}<\sigma_{ij}+\Delta^0_{ij}+x~\Delta^1_{ij} \\
0&  \sigma_{ij}+\Delta^0_{ij}+x~\Delta^1_{ij}<r_{ij} 
\end{cases}
\end{equation}
where $\Delta^0_{ij}=\lambda_0 \sigma_{ij}$ and $U_0$ are the width and depth, respectively, of the square well, and $\Delta^1_{ij}=\lambda_1 \sigma_{ij}$ and $U_1$ are the corresponding parameters for the square shoulder. Model (i) has $x=0$, while model (ii) has $x=1$, and in both cases temperature, $T$, is expressed in reduced units of $U_0$ with Boltzmann constant, $k_\mathrm{B}$, set to unity. Hence, model (i) has a single tuning parameter, $\lambda_0$, while model (ii) has three: $\lambda_0$, $\lambda_1$, and $U_1$. We consider the dynamics of these systems at constant $N$, volume $V$ and $T$ using a Monte Carlo dynamics that only consists of $N$ single-particle translations per unit time, $t$. These translation are taken along a vector randomly drawn within a three-dimensional cube of side $\delta \ell$, such that the relaxation time is minimum at a packing fraction close to the dynamical transition. The results of such Monte Carlo dynamics are known to be similar to those of other \emph{local} dynamics in the dense fluid regime which is the regime of interest for this work\c{ludo_mc,kob_mc,dyn_mc}.  

Equilibration of the initial system is ensured by running Monte Carlo dynamics for at least ten structural relaxation times, $\tau_\alpha$, defined from the characteristic decay, $Q(\tau_{\alpha}) \equiv e^{-1}$, of the self-part of the particle-scale overlap function
\be
Q(t)=\frac{1}{N}\sum_{i=0}^{N} \Theta(a-|r_i(t)-r_i(0)|), 
\ee
where $\Theta$ is a step function and $a = 0.3 {\sigma}_{\mathrm{B}}$ is a microscopic length chosen to be close to the typical particle cage size.  This function therefore represents the fraction of particles having moved a distance smaller than $a$ by time $t$. 

\begin{figure}
\includegraphics[width=0.98\columnwidth]{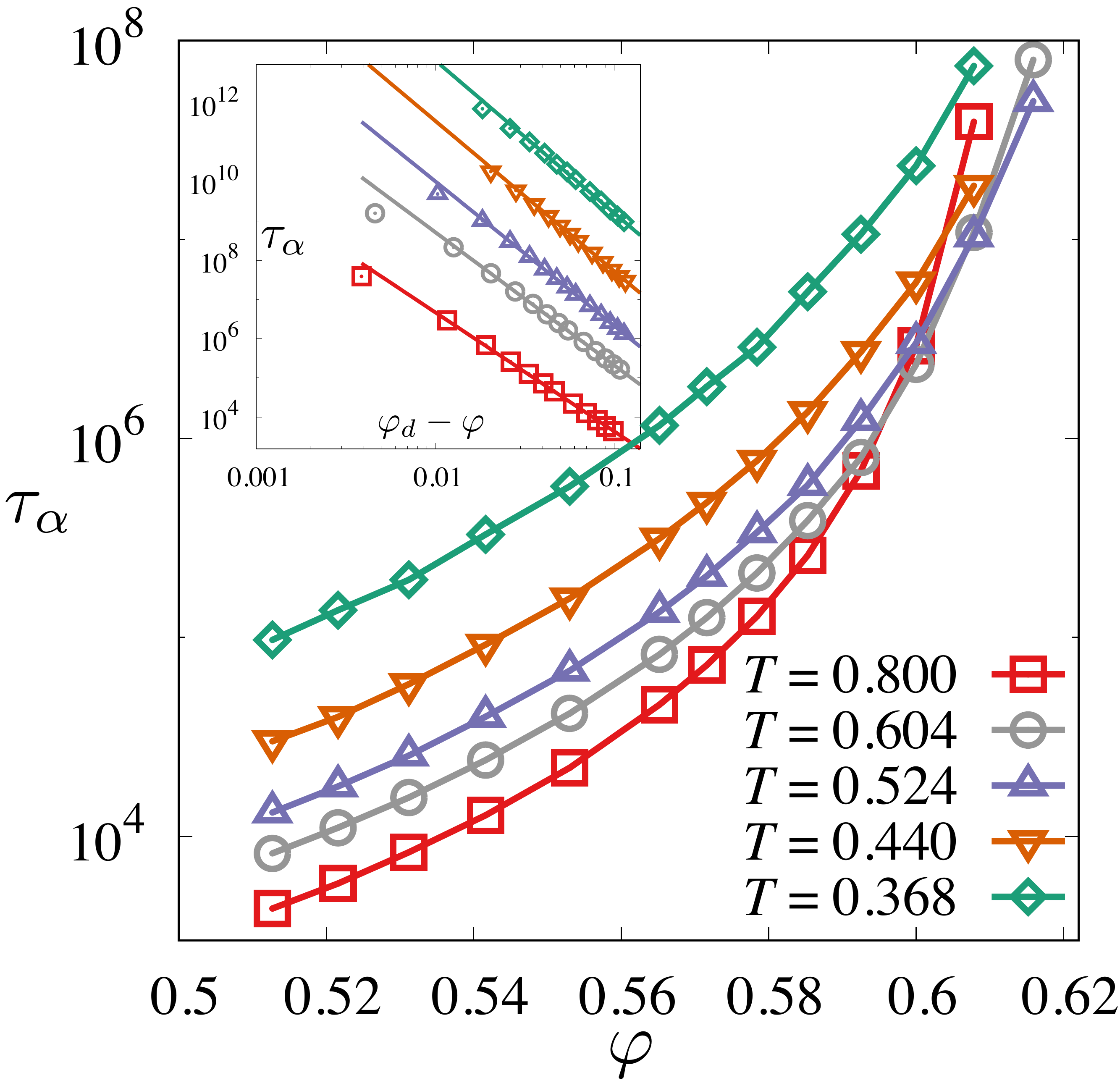}
\caption{The relaxation time $\tau_\alpha$ as a function of the packing fraction $\varphi$ for different temperatures. Inset: dynamical transition densities, $\varphi_\mathrm{d}(T)$, estimated by fitting the structural relaxation times to a power-law $\tau_\alpha(\varphi;T)=A(\varphi_\mathrm{d}(T)-\varphi)^{-\gamma}$. Results here are given for a model with $\lambda_0=0.019$, $\lambda_1=2.5$, $U_1=0.10$. Deviations from the power-law as $\varphi\rightarrow\varphi_{\rm d}$ are due to activated cage escapes. For visual clarity, the vertical scale  for $T=0.800, 0.604, 0.524, 0.440$ and $0.368$ has been multiplied by $10^{0},10^{1}, 10^2, 10^3,$ and $10^4$, respectively.}
\label{gamma}
\end{figure}
\begin{figure}
\includegraphics[width=\columnwidth]{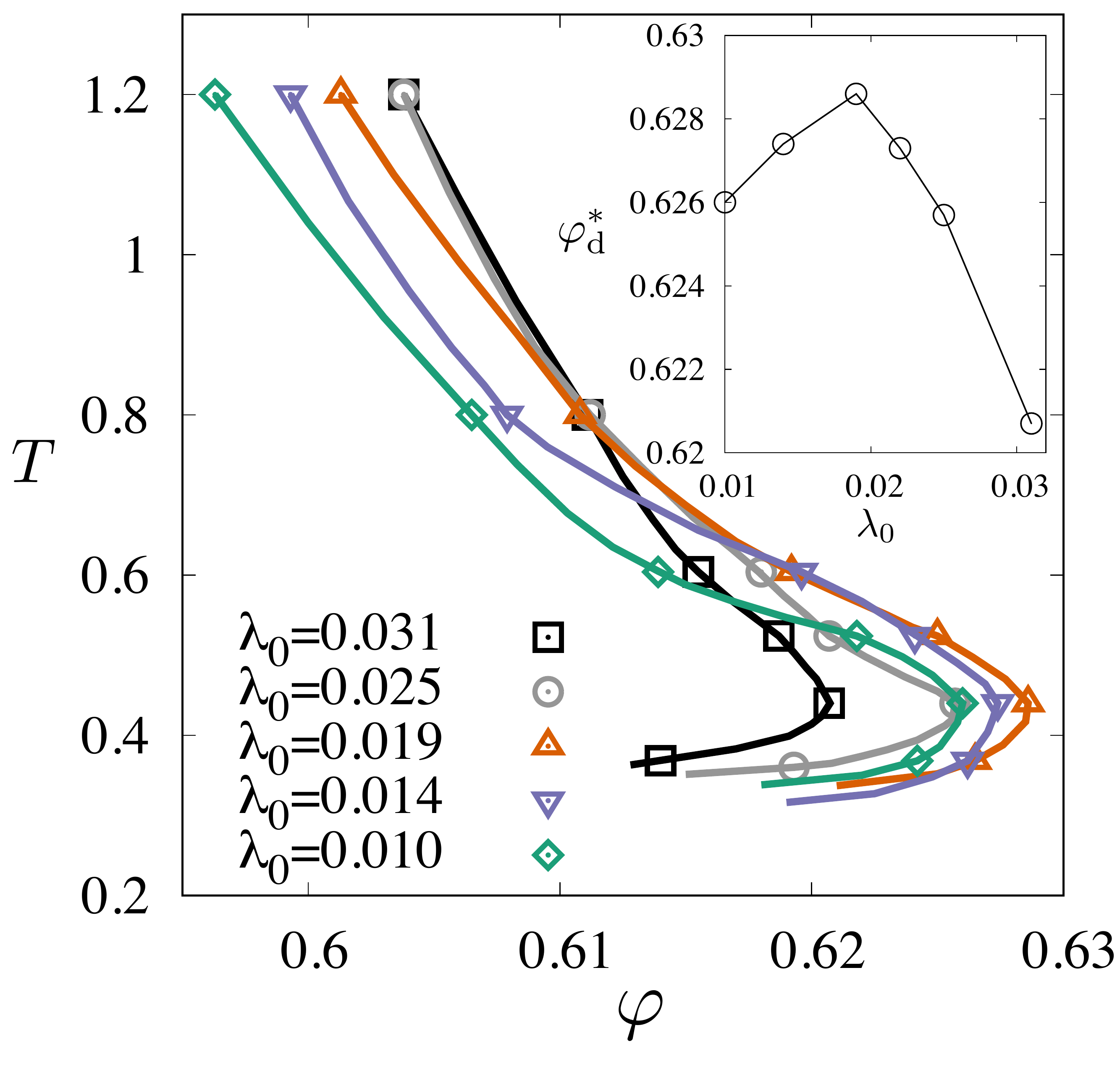}
\caption{Dynamical diagram for fluids of spheres interacting via a square-well attraction of different well widths $\lambda_0$. Inset: the maximum fluid packing fraction $\varphi_{\rm d}^{\mathrm{\ast}}$ accessible from the liquid side for different ranges of attraction $\lambda_0$.}
\label{SW}
\end{figure}

The equilibrium $Q(t)$ for the liquid is averaged over the trajectory that begins after equilibration. Typical plots for the relaxation time as a function of the packing fraction are shown in \f{gamma} for different temperatures. At fixed $T$, $\tau_{\alpha}(\varphi;T)$ is used to estimate the (avoided) dynamical transition density, $\varphi_\mathrm{d}(T)$, by fitting to the critical scaling form, $\tau_\alpha(\varphi;T)=A(\varphi_\mathrm{d}(T)-\varphi)^{-\gamma}$-- see inset of \f{gamma}. Because of the presence of activated processes in finite dimensions, this power-law scaling persists for at most a couple of decades\c{patrick_review}, but this range suffices to provide a fairly robust estimate of $\varphi_\mathrm{d}$. 
Estimation of $\varphi_{\rm d}(T)$ provides the dynamical diagram in the $\varphi$-$T$ plane. 

\section{Results and Discussion} 
\label{sec:results}
We first tune the interaction range of the simple SW system in order to maximize the depth of the fluid pocket. To the best of our knowledge this optimization had not been previously attempted in simulations. Most previous studies considered models with $\lambda_0=0.03$, following the MCT prediction for the existence of an anomalous glassy regime for that interaction range. The dynamical diagrams for different $\lambda_0$ around 3\% are shown in \f{SW}; the dynamical reentrance of the liquid is clearly visible. The maximum accessible liquid density, $\varphi^{\ast}_{\rm d}$, is however, not attained with $\lambda_0=0.03$, but rather with one of $\lambda^{\ast}_0 \approx 0.019$. Although our result for $\lambda^{\ast}_0\approx 0.019$ is in the vicinity of the infinite-dimensional theoretical prediction for this optimization ($\lambda^{\ast}_0 \approx 0.029$)\c{francesco2018}, it is nonetheless significantly different from it. Because the intricate liquid structure of finite-dimensional systems is neglected in the analytical study, this discrepancy is not particularly surprising. In three dimensions, the nearest-neighbor shell structure is indeed much tighter than what is theoretically assumed. A possible explanation for the discrepancy is therefore that a smaller attraction range suffices in simulations to obtain an energetic stabilization comparable to what is expected in the $d\rightarrow\infty$ limit.
\begin{figure}
\includegraphics[width=\columnwidth]{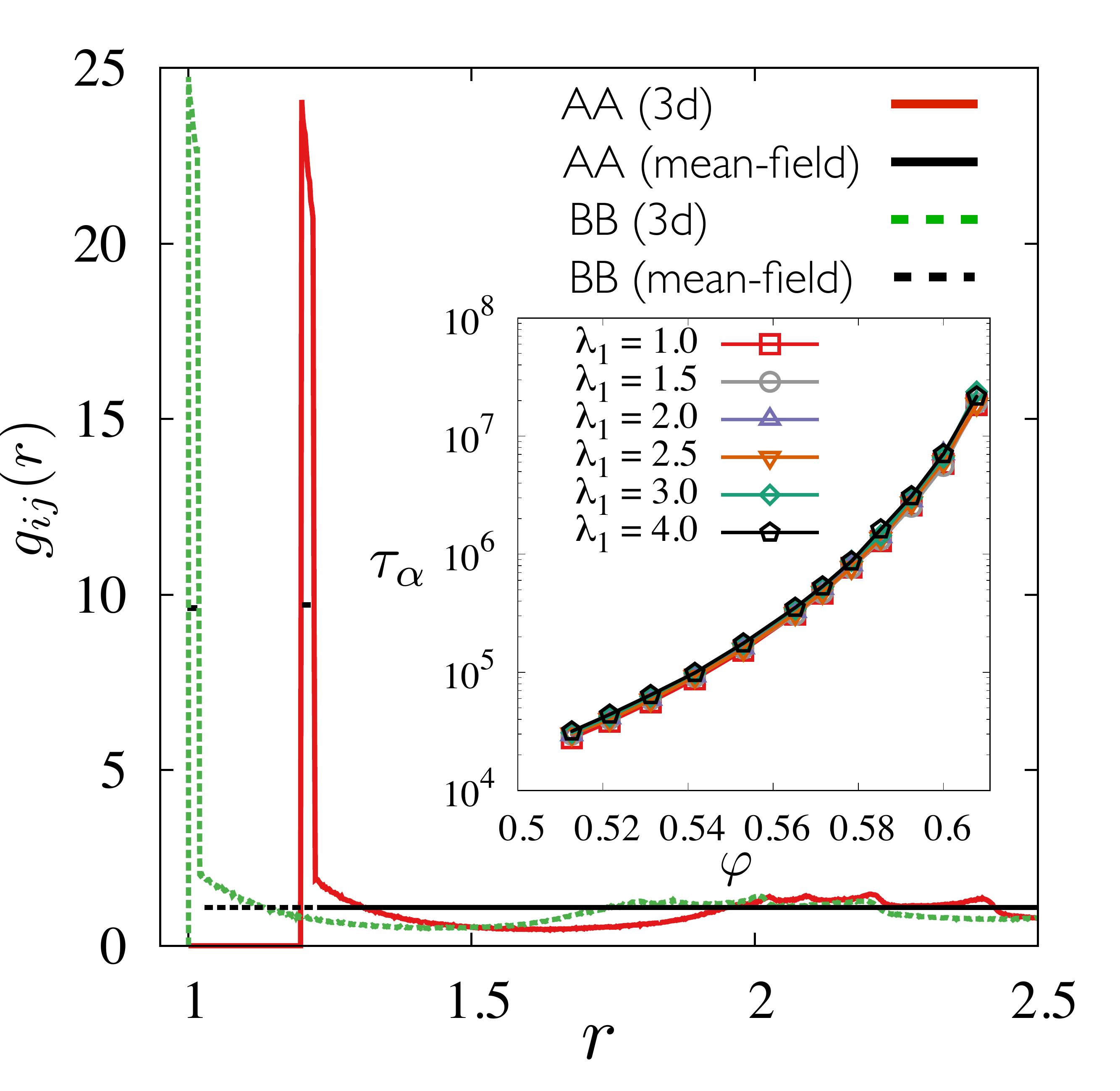}
\caption{Partial pair correlation function $g_{ij}(r)$ for A and B particles. The liquid shell structure is much stronger in $d=3$ than in the $d\rightarrow\infty$ limit. Inset: the evolution of $\tau_\alpha$ with $\varphi$ is remarkably insensitive to the choice of $\lambda_1$, but very small deviations can be seen when $\lambda_1 > 2.5$.}
\label{g_of_r}
\end{figure}
\begin{figure}
\includegraphics[width=\columnwidth]{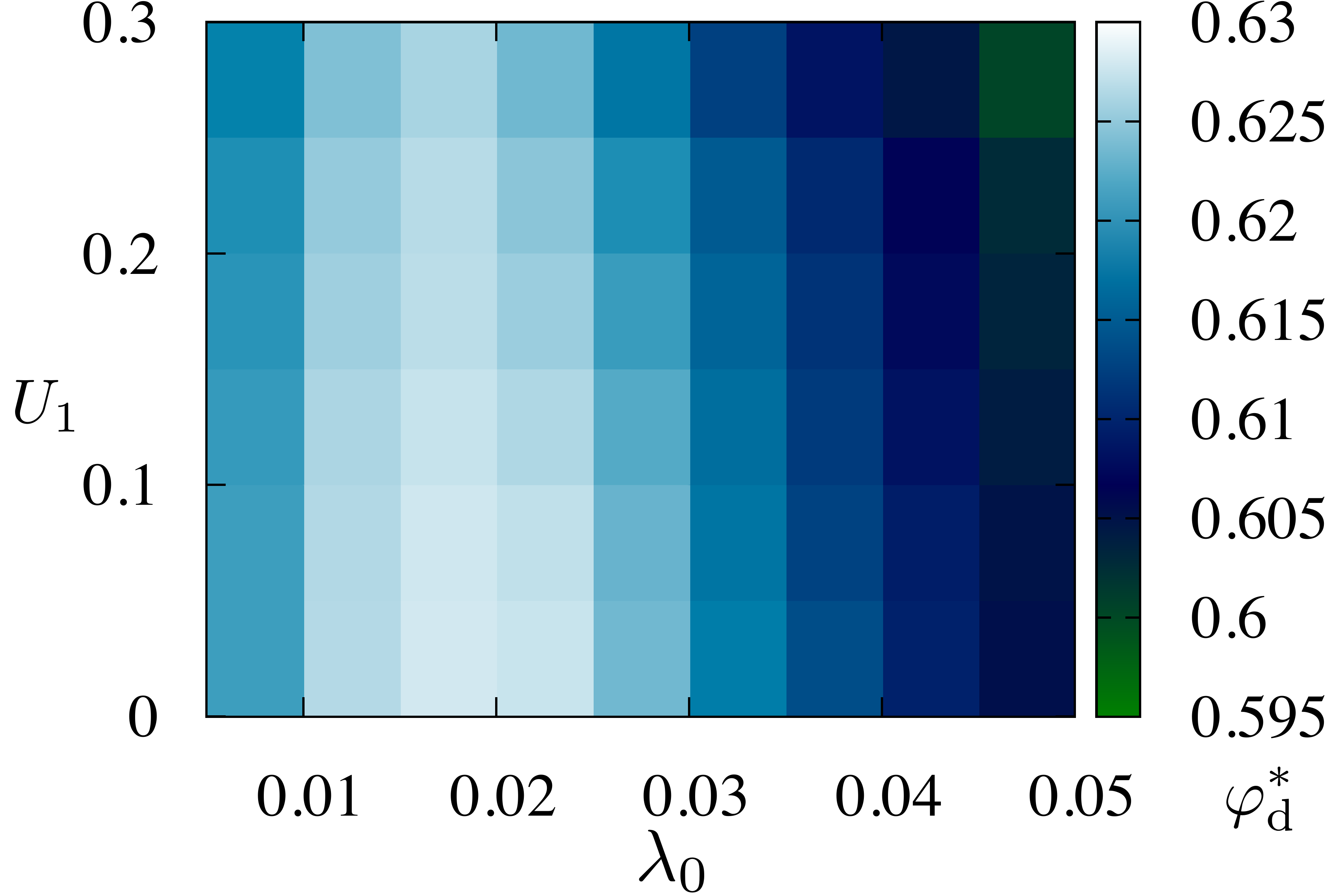}
\caption{Maximal fluid packing fraction $\varphi_{\rm d}^{\mathrm{\ast}}$ for the SW+SS system in the parameter space of $U_1$ and $\lambda_0$ for $\lambda_1=2.5$. The line $U_1=0$ corresponds to the SW system. This plot reveals that longer-ranged repulsion does not significantly push $\varphi_\mathrm{d}^{\ast}$ to higher densities in three dimensions, but nonetheless gives rise to a parameter pocket of enhanced reentrance around $\lambda_0\approx 0.019$, away from the $U_1=0$ axis.}
\label{color}
\end{figure}

We next explore whether adding a suitably tuned repulsive part to the potential can further delay the dynamical arrest. In this case, three parameters are to be optimized: $\lambda_0$, $\lambda_1$, and $U_1$. We expect the three-dimensional parameter space ($\lambda_0$, $\lambda_1$, $U_1$) for the SW+SS system to be simple with a single minimum (corresponding to the densest liquid configuration) connected to the minimum of the SW system ($U_1=0, \lambda_1=0$) through a path without large barriers. To nonetheless ensure that our optimization does not miss its target, we explore a wide range of parameter values. We search for an optimum over $\lambda_0\in(0.010,0.060)$, $\lambda_1\in(0.5,5.0)$, and $U_1\in(0.0, 0.40)$ by gridding the parameter space, and compute $\varphi_\mathrm{d}$ for a few temperatures around the reentrance regime in the dynamical diagram for each grid point to estimate $\varphi_\mathrm{d}^{\ast}$. From this scheme we identified the set of parameters that pushes the dynamical transition furthest as $\lambda^{\ast}_0 = 0.019 \pm 0.004$, $U^{\ast}_1 \leq 0.10$, and $0.8 \leq \lambda^{\ast}_1 \leq 3.0$. All directions away from this optimum lead to lower or comparable values of $\varphi_{\rm d}^{\ast}$. 
The optimal parameters identified are in qualitative agreement with the theoretical prediction that the repulsion should be much weaker and longer ranged than the attraction and that the attraction range does not markedly broaden in going from a SW to a SW+SS model. In our case, however, the attraction range barely changes, while the theoretical prediction has $\lambda^{\ast}_0$ increase from $0.029 \to 0.054$. Here again, the tightness of the finite-dimensional neighbor shell is likely at play. 
\begin{figure}
\includegraphics[width=\columnwidth]{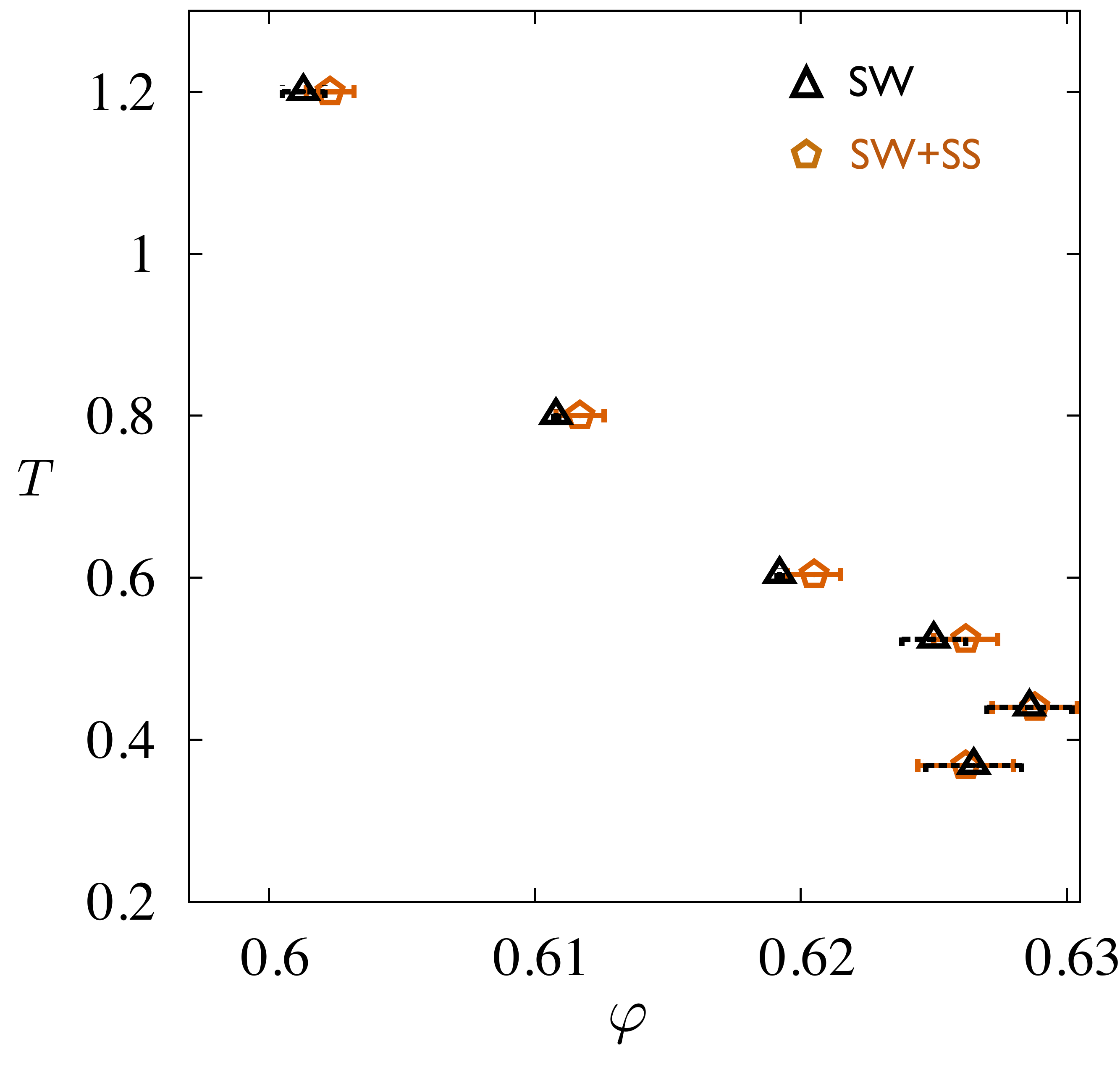}
\caption{Dynamical diagram for the SW system with $\lambda_0=0.019$ and the SW+SS system with optimized parameters ($\lambda_0=0.019$, $\lambda_1=2.5$, $U_1=0.10$). Adding longer-ranged repulsion does not significantly push $\varphi_\mathrm{d}^{\ast}$ to higher densities in these models.}
\label{SW_SS}
\end{figure}

A more significant difference is that while the dynamics (and thus $\varphi_\mathrm{d}$) is somewhat sensitive to the repulsion strength $U_1$, its dependence on the repulsion range $\lambda_1$ is much weaker over the parameter window considered. The dynamics of all models with $\lambda_1\in [0.5,2.5]$ indeed roughly coincides (see \f{g_of_r} inset).  Once more, the pronounced shell structure of three-dimensional dense fluid is likely at play. While in the $d\rightarrow\infty$ limit, $g(r) = \exp(-\beta V_{ij} (r))$, and hence self-solvation can be strongly impacted by the interaction potential; in three dimensions, the influence of the hard core-repulsion is felt much more strongly (see \f{g_of_r}). As a result, adding a weak repulsive contribution to the interaction potential results in a much weaker dynamical effect. For a repulsion range that falls within the intershell depletion regime, no notable effect on the dynamics are thus observed.

Given the relative insensitivity of the optimization to $\lambda_1$, we can concentrate on the two-dimensional parameter space, $\lambda_0-U_1$, for $\varphi_\mathrm{d}^{\ast}$. \f{color} shows the maximum fluid packing fraction $\varphi_{\rm d}^{\ast}$ in the space of $U_0$ and $\lambda_1$ where $\lambda_1=2.5$. Interestingly, the optimization landscape is relatively flat along $U_1$. The SW optimum is therefore connected by a fairly soft mode to the SW+SS optimum. Along $\lambda_1$, by contrast, $\varphi_{\rm d}^{\ast}$ changes much more rapidly. This landscape projection is therefore consistent with the above discussion. 

The resulting dynamical diagrams for the SW and the SW+SS optima are compared in \f{SW_SS}. The results show that the corresponding $\varphi_{\rm d} (T)$ values are not significantly different (within numerical uncertainty) from one another. If any enhancement of the reentrance pocket is present, it is therefore much smaller than the $3\%$, predicted by the $d = \infty$ calculation.

\section{Conclusion}
\label{sec:concl}

Motivated  by a recent mean-field prediction that the dynamically sluggish fluid regime for models with SALR interactions can be pushed to higher densities than for models with purely short-ranged attraction, we have performed extensive Monte Carlo simulations of a family of SW and SW+SS models. Our exploration of model parameters did not identify (within numerical uncertainty) any SALR model that pushes the dynamical transition significantly beyond the densest packing achievable by only short-ranged attraction. We did, however, identify a branch of parameters over which the optimum extends. This nontrivial feature could be a finite dimensional echo of the $d\to \infty$ prediction. The theoretical prediction that further tuning the interaction potential could engender additional (smaller) gains in $\varphi_{\mathrm{d}}^\ast$~\cite{francesco2018} is nevertheless unlikely to be verifiable in three-dimensional systems. 

Data associated with this work are available from the Duke Digital Repository at ``\textit{will be added}".

\begin{acknowledgements}
This paper is dedicated to the late Bob Behringer, who has always been warm, wise and supportive to this junior colleague (PC). He will be sorely missed. We acknowledge funding from the Simons Foundation (Grant \# 454937 to PC) and computer time of Duke Compute Cluster
(DCC) and Extreme Science and Engineering Discovery Environment (XSEDE), supported by National Science Foundation grant number ACI-1548562. 
\end{acknowledgements}

%

\end{document}